\documentclass[prl,aps,showpacs,floatfix]{revtex4}

\usepackage{epsfig}
\usepackage{amssymb}
\usepackage{amsmath}
\usepackage{amsfonts}
\usepackage{epstopdf}
\usepackage{verbatim}
\usepackage[normalem]{ulem}
\usepackage{lipsum}

\usepackage{verbatim}

\usepackage{float}

\usepackage{graphicx}
\usepackage{color}
\usepackage{subcaption}

\begin{document}
\title {
Novel Energy Scale  in the Interacting 2D Electron System Evidenced  from
Transport and Thermodynamic
Measurements
}

\author{L.\,A.~Morgun$^{1,2}$ , A.\,Yu.~Kuntsevich$^{1,2}$, and V.\,M.~Pudalov$^{1,3}$\thanks{e-mail:
pudalov@lebedev.ru}}

\address{$^1$ P.\,N.~Lebedev Physical Institute, 119991 Moscow, Russia\\
$^2$ Moscow Institute  of Physics and Technology, Moscow 141700, Russia\\
$^3$ National Research University Higher School of Economics, Moscow 101000, Russia}

\begin{abstract}
We study how the non-Fermi-liquid two-phase state reveals itself in
transport properties of high-mobility Si-MOSFETs.
 We have found features in zero-field transport, magnetotransport, and thermodynamic spin magnetization in a 2D
 correlated electron system
 that may be directly related with the two-phase state. The features manifest above a  density
dependent temperature $T^*$ that represents  a novel high-energy scale, apart from the Fermi energy.
More specifically,  in magnetoconductivity, we found a sharp onset of the novel regime $\delta \sigma(B,T)
\propto
(B/T)^2$ above a density-dependent temperature $T_{\rm kink}(n)$,
a high-energy behavior that ``mimics'' the low-temperature diffusive interaction regime.  The zero-field
resistivity temperature dependence exhibits an inflection point  $T_{\rm infl}(n)$.
 In thermodynamic magnetization, the weak-field spin susceptibility per electron, $\partial \chi /\partial n$
 changes
 sign at $T_{dM/dn}(n)$.
  All three notable  temperatures, $T_{\rm kink}$, $T_{\rm infl}$, and $T_{d M/ d n}$, behave critically $\propto
  (n-n_c)$,  are close to each other,  and are intrinsic to high-mobility samples solely; we therefore associate
  them with an energy scale $T^*$ caused by interactions in the 2DE system.

\end{abstract}
\pacs{71.30.+h, 73.40.Qv, 71.27.+a}
\maketitle


\section{Introduction}
Two-dimensional  (2D) interacting low density carrier systems in the past two decades attracted considerable
interest
\cite{kravreview_2001, pud_granada_2004, dolgopolov_UFN_2003, shashkin_UFN_2005}, demonstrating fascinating
electron-electron interaction effects, such as
metallic temperature dependence of resistivity
\cite{krav_PRB_1994, hanein_1998, papadakis_1998},  metal-insulator
transition (MIT) \cite{krav_PRB_1994, krav_PRB_1995, amp_2001, kravreview_2001, krav-sarachik_Reprog_2004}, strong
positive magnetoresistance  (MR) in parallel field
\cite{simonian_PRL_1997, pudalov-MR_JETPL_1997, pudalov-MR_PhysicaB_1998,
 yoon_PRL_2000, pudalov-aniso_PRL_2002, lai-Si-SiGe_PRB_2005, lu_PRB_2008, vitkalov-MR-insulator_PRB_2005}, strong
renormalization of the effective mass and spin susceptibility \cite{okamoto_PRL_1999, pudalov-gm_PRL_2002,
zhu_PRL_2003, tutuc_SdH_PRL_2003,  pud_granada_2004, shashkin-m_PRB_2002,  clarke_NatPhys_2008}, etc.

Far away from the critical MIT density $n_c$, in the well ``metallic regime,'' these effects are
explained within  the framework of the Fermi liquid theory -- either in terms of interaction quantum corrections
(IC)
\cite{ZNA-R(T)_PRB_2001, ZNA_BPar_2001}, or
temperature-dependent screening of the disorder  potential  \cite{gold-dolgopolov_PRB_1986, das-hwang,
das-hwang_PRB_2013, das-hwang_PRB_2014, g-d-MR_JETPL_2000}.
Both theoretical approaches so far are used to treat the experimental  data on transport, and the former one --
also
  to determine the Fermi liquid coupling constants from fitting the transport and magnetotransport data to the IC
  theory.
In the close vicinity of the critical region,  conduction is treated within the renormalization group
\cite{finkelstein_1984,
castellani_PRB_1984, CCL_PRB_1998, punnoose_PRL_2002, pf_science_2005},
or the Wigner-Mott approach  \cite{camjayi-dobro_natphys_2008, dobro-kotliar_WM_PRB_2012}.

On the other side,  a number of theories predicts breakdown of the uniform paramagnetic 2D Fermi liquid state due
to
instability in the spin or charge channel, developing as
interaction strength increases   \cite{dharma_2003, narozhny_2000, khodel_2005, spivak, sushkov}.
 However,   how the potential instabilities  reveal themselves in charge
transport remains an almost unexplored question.

\subsection{On the spin polarization of the  2D electron system }
Spin fluctuations are believed to play an important role in the 2DE system, especially near the apparent
metal-insulator transition.
Ferromagnetic instabilities result from the interplay of the electronic interactions and the Pauli
principle. The interaction energy can be minimized when the  fermion antisymmetry requirement is satisfied by the
spatial wave function resulting in the alignment of spins and a large ground-state spin magnetization.
In clean  metals, the long-range part of the Coulomb interaction is screened, whereas its short-range part  leads
to
strong correlations of the electron liquid.
This short-range part of the interaction leads to ferromagnetic (Stoner) instability at sufficiently
large values of the interaction strength.
Initial numerical  quantum Monte Carlo  calculations   \cite{tanatar-ceperley_1989}  did not reveal a difference
in
energy between the polarized and unpolarized fluid phases at the crystallization transition.
 From diffusion Monte Carlo  calculations \cite{varsano_epl_2001},
 no evidence was found for the stability of a {\em partially} spin-polarized fluid phase in 2D systems.

The valley degree of freedom has qualitative effects on the 2DEG properties,
making the fully spin-polarized fluid unstable \cite{conti_epl_1996, marchi_prb_2009}, at
variance with the one-valley 2DE system. This conclusion directly refers to  the two-valley electron system in
(100) Si-MOS samples.
The DMC calculations \cite{marchi_prb_2009} confirm the absence of a transition from the paramagnetic to the fully
spin-polarized fluid in the two-valley symmetric system.
Moreover, in the whole density
range, where the fluid is stable, there is no evidence for the stability of a state with partial spin polarization
\cite{varsano_epl_2001, marchi_prb_2009, tanatar-ceperley_1989, attaccalite_PRL_2002, dharma_2003}.

\subsection{Spin polarization of the  spatially confined 2DE system}
In Ref.~\cite{benenti_prl_2001},  the ground-state magnetization was numerically studied for clusters of
interacting
electrons
in two dimensions in the regime where the single-particle wave functions are localized by disorder.  It is found
that
the Coulomb interaction leads to a spontaneous ground-state magnetization.  The magnetization is suppressed when
the
single-particle states become delocalized.
The stability of the minimum spin ground state in a quantum dot was
analyzed in Ref.~\cite{prus_PRB_1996}. Within perturbation
theory, the effective interaction strength is enhanced by the
presence of disorder, leading to a ferromagnetic instability already below the Stoner threshold
 \cite{andreev-kamenev_prl_1998}.
Observations of the spin polarization for a few electrons system  confined in quantum dots were  reported in
several
experiments
\cite{ghosh_PRL_2004, rogge_PRL_2010} and are considered as evidence of interaction-induced collective spin
polarization transition.

\subsection{Experimental situation}

For the infinite 2D electron  system, extensive experimental search has been undertaken and the results are
contradictory
enough.
The respective reviews may be found in Refs.~\cite{pud_granada_2004, dolgopolov_UFN_2003, shashkin_UFN_2005,
sarachik_JSPS_2003, pudalov_Varenna_2004, clarke_NatPhys_2008}.
The experimental results may be summarized as follows.
From experiments with low perpendicular fields, the spin susceptibility of itinerate electrons, determined from
quantum oscillations,  remains finite down to the critical density of the 2D metal-insulator transition, $n=n_c$
\cite{okamoto_PRL_1999, pudalov-gm_PRL_2002, zhu_PRL_2003, tutuc_SdH_PRL_2003, pud_granada_2004}. In particular,
the spin susceptibility was measured in GaAs/AlGaAs superlattices
\cite{zhu_PRL_2003}, with electron densities as low as $1.7\times 10^9$cm$^{-2}$ and
no polarization transition was observed.

In contrast, the susceptibility and effective mass determined with (100)Si-MOS samples from in-plane field ($B>T$)
magnetotransport \cite{vitkalov_PRL_2001, shashkin_PRL_2001} and temperature-dependent transport
\cite{shashkin-m_PRB_2002}, were
reported to diverge;  based on these data, the authors concluded on the ferromagnetic instability of itinerant
electrons in 2DE system.
In similar experiments \cite{pudalov-aniso_PRL_2002, pudalov_R(T)_PRL_2003, klimov_PRB_2008, lu_PRB_2008,
lai-Si-SiGe_PRB_2005, yoon_PRL_2000}, however,  the opposite conclusion was achieved, that is, the ferromagnetic
instability does not
occur and the spin susceptibility
remains finite down to the lowest accessible density, e.g.,  down to $n=0.3\times 10^{11}$cm$^{-2}$ in Si/SiGe
quantum
wells in Ref.~\cite{lu_PRB_2008}. In measurements of the weak localization
\cite{proskuryakov_PRL_2001} and quantum oscillations in a  weak perpendicular field \cite{pud_granada_2004}, a
Fermi-liquid type behavior was found  with no signatures of the spin polarization of itinerant electrons.

Eventually, the thermodynamic spin magnetization measurements  performed in a weak field \cite{teneh_PRL_2012}
have
clarified the reason of the
contradiction: the 2D interacting electron system experiences a transition from Fermi
liquid to the two-phase state,  that hampered interpretation of the data.
The main result of the thermodynamic weak field studies
is the observation of ``spin-droplets'' -- spin-polarized collective electron states with a total spin of the order
of
two
\cite{teneh_PRL_2012}. These easily polarized ``nanomagnets'' exist as a minority phase on the background of the
majority Fermi liquid phase even though the density and the dimensionless conductance  are high, $k_F l \sim 100
$;
the latter is commonly considered as a criterion of the well-defined Fermi liquid state.

\subsection{Motivation}
In this paper we study how the non Fermi-liquid two-phase state reveals itself in
  magnetotransport and zero-field transport. We
report results  of the transport and magnetotransport  measurements with a 2D correlated
electron system, which reveal the existence of a
characteristic energy scale $T^*$ that is smaller than the
Fermi temperature $T_F$, but much bigger than $1/\tau$
 (we set throughout the paper   $\hbar, k_B, e  = 1$).
The same energy scale is found in our earlier magnetization measurements.
Obviously, no such large energy scale may exist in the Fermi liquid.
In magnetoconductivity $\sigma(B_\parallel)$, we found a sharp onset of the  novel regime $\delta
\sigma(B,T) \propto (B/T)^2$ above a density-dependent $T_{\rm kink}(n)$, the high-energy behavior that ``mimics''
the low-temperature diffusive interaction
regime \cite{ZNA_BPar_2001}.
 In zero-field transport,  there is an  inflection point $T_{\rm infl}(n)$  on the
  resistivity temperature dependence. We found that the two remarkable temperatures
are close to each other and close to the temperature $T_{dM/dn}$ for which the spin susceptibility per electron
$\partial \chi
/\partial n$ (and $\partial M/\partial n$) changes sign (the phenomenon  reported earlier in Ref.
\cite{teneh_PRL_2012}). All
three notable  temperatures, $T_{\rm kink} \approx T_{\rm infl}\approx T_{dM /dn}$,  behave critically $\propto
(n-n_c)$, and are
intrinsic to high-mobility samples only;
we therefore associate them with an
energy scale $T^*$ caused by interactions in the 2DE system. Our studies  do not address critical behavior at MIT
and in the $T=0$ limit, rather,  we focus  on the high-density regime,  away from the critical density of the 2D
MIT, and on the high-temperature regime where resistivity exhibits strong growth with temperature.

\section{ Experimental}
The ac measurements (5 to 17\,Hz) of resistivity were performed using a four-probe lock-in technique
 in magnetic fields up to $\pm 7$\,T. The  range of temperatures,   0.4-20\,K, was chosen so as to ensure the
 absence of the shunting conduction of bulk Si at
 the highest temperatures, and, on the low-temperature side,
  to exceed the valley splitting and intervalley scattering rate  \cite{valley_scattering}.
   Measurements were performed with three  ``high-mobility'' samples, Si-2, Si-63, and Si-4  ($\mu^{\rm
    peak} = 3$, 2.5, and 0.95\,m$^2$/Vs), and, for comparison, with two ``low-mobility'' samples
 Si-40 and  Si-46 ($\mu^{\rm max} \approx 0.2$ and 0.1m$^2$/Vs). Their transport features are described further.
All samples had
  $\approx 190 \pm 5$\,nm gate oxide   thickness, and were
  lithographically defined as rectangular  Hall bars,  $0.8\times5$\,mm$^2$.
The magnetoconductivity measurements were performed similar to
Ref.~\cite{knyazev_JETPL_2006},
but in a much wider domain of densities and temperatures, from far above the MIT critical density ($n\gg n_c$) and
in
the well-conducting regime $k_Fl \gg 1$ down to
low densities $n\gtrsim n_c$.

By rotating the sample with a step motor, we  aligned the magnetic field in the 2D plane to within $1^\prime$
accuracy,
using the  weak localization magnetoresistance as a sensor of the perpendicular field component.
The carrier density $n$ was varied by the gate voltage $V_g$ in the range $(0.9-10)\times 10^{11}$\,cm$^{-2}$. The
linear
$n(V_g)$ dependence was determined from the quantum oscillations period measured in the
perpendicular field orientation during the same cooldown.

\section{Experimental results}
\subsection{In-plane field magnetoconductivity}

The lowest-order  variations of the conductivity (as well as resistivity) with a weak in-plane field  $g\mu_B B<T
\ll
T_F$ at a fixed temperature $T$
are  parabolic. This follows from the
symmetry arguments, as well as from the interaction correction theory and the screening theory:
\begin{eqnarray}
\sigma &=& \sigma_0 - a_\sigma  B^2 + {\cal{O}}\left(B^2\right) \nonumber \\
\rho &=& \rho_0 + a_\rho B^2 + {\cal{O}}(B^2),
\label{eq:sigma}
\end{eqnarray}
 where  $g\mu_B B$ is considered to be small as compared with either $T$,  $(T^2\tau)$, or $T_F$,
  and by definition
\begin{eqnarray}
a_\sigma &\equiv & \left. -\frac{1}{2} \partial^2\sigma/\partial B^2 \right|_{B=0} \nonumber \\
a_\rho &\equiv & \left. \frac{1}{2} \partial^2\rho/\partial B^2 \right|_{B=0} \nonumber.
\end{eqnarray}

\begin{figure}
\begin{center}
\includegraphics[width=0.49\linewidth]{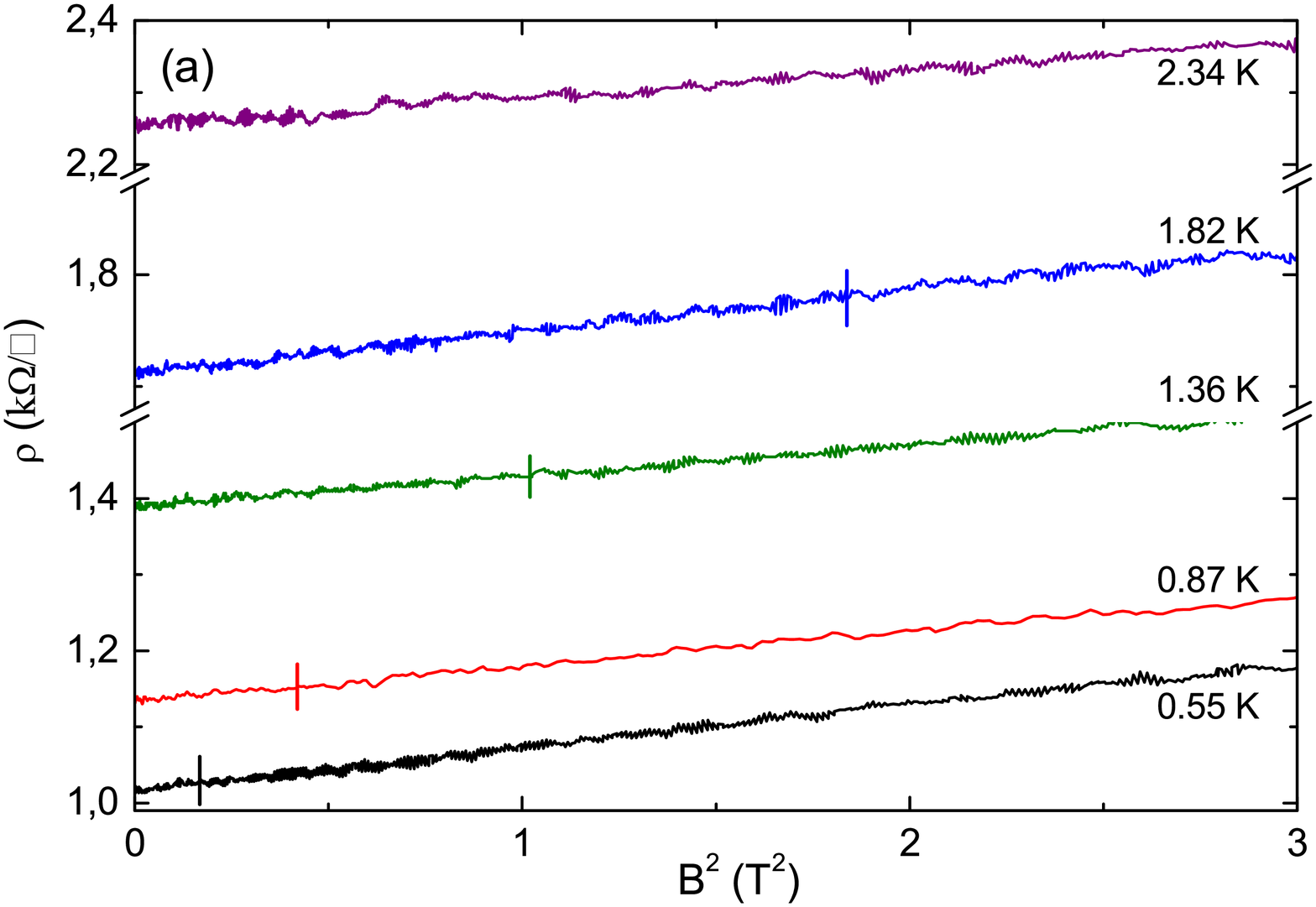}
\includegraphics[width=0.49\linewidth]{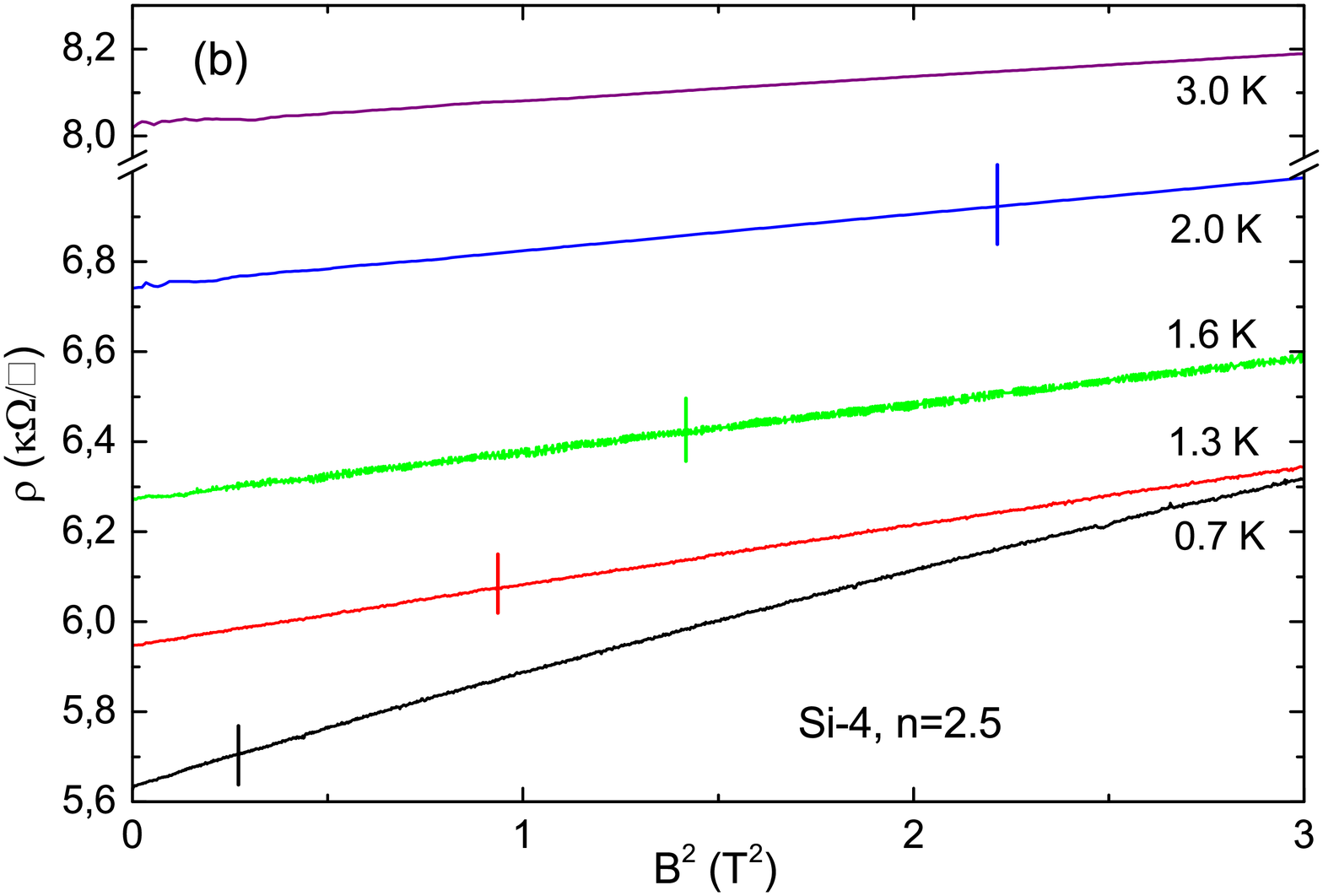}
\caption{Magnetic field dependence of the resistivity
(a) for sample Si-2 at  five temperatures: 0.55, 0.87, 1.36, 1.82, and 2.34\,K,  electron density is $2\times
10^{11}$cm$^{-2}$, and
(b) for sample Si-4 at five temperatures: 0.7, 1.3, 1.6, 2.0, and 3.0\,K, electron density is $2.5\times
10^{11}$cm$^{-2}$.
Vertical ticks mark  the $g\mu_B B=T$ field.
}
\label{fig:Si4-rho(B2)}
\end{center}
\end{figure}

In the experimental data, a purely parabolic $\rho(B)\propto B^2$ dependence was found to extend with high
accuracy
even far above
the range of low fields ($g\mu_B B<T$) (see  Fig.~\ref{fig:Si4-rho(B2)}). For this reason, we quantified the
magnetoconductivity
using the prefactor $a_\sigma(T,n)$.
For example, the higher order-in-$(g\mu_B B/T)$ terms in Eq.~(\ref{eq:sigma})
were less than  0.1\%  (relative to the $B^2$-term) even at $g\mu_B B/T = 6.5$, and could be safely neglected
therefore for $g\mu_B B\ll T$.

Consider the relation between $a_\sigma$ and the experimentally measured magnetoresistance (MR) $\rho(B)$.
In purely parallel magnetic field  $\sigma = 1/\rho$. Taking the second derivative  from both sides of equation
(\ref{eq:sigma})
and recalling that
$\left. \left(\partial \rho/\partial B\right) \right|_{B=0} = 0$, we obtain
\begin{equation}
a_\sigma = \left[\frac{1}{2 \rho^2} \frac{\partial^2 \rho}{\partial B^2} - \frac{1}{\rho^3} \left(\frac{\partial
\rho}{\partial B}
\right)^2\right]_{B=0} = \frac{1}{2 \rho^2} \frac{\partial^2 \rho}{\partial B^2}.
\end{equation}
Following the latter relation, from the experimentally measured magnetoresistivity,
we determined the magnetoconductivity prefactor,
which is analyzed below versus $T$ for various densities.

The variations of the conductivity with a weak in-plane field at a fixed temperature are low, $\leq 5\%$, in the
selected
range of fields
$g\mu_B B<T$ \cite{remark_B/T} (see  Figs.~\ref{fig:Si4-rho(B2)}).
This smallness favors comparison of the data with theory of interaction corrections (IC),
which makes firm predictions specifically for magnetoconductivity (MC)
 and suggests a clear physical picture behind it \cite{ZNA_BPar_2001}.

In the spirit of the IC theory, the
temperature variation of the conductivity of the 2DE system is described by the  interference
and e-e interaction corrections \cite{ZNA-R(T)_PRB_2001}
$$
\Delta\sigma(T) = \Delta\sigma_C(T) +n_T\Delta\sigma_T(T) +O\left(\frac{1}{k_Fl}\right).
$$
Here, the first term combines both single-particle interference and  interaction corrections in the singlet
channel,
and the second term is the interaction corrections in the triplet channels whose number depends on the valley
degeneracy,
$n_T=4g_v^2 -1$ \cite{punnoose_PRL_2002}, and $k_F l$ is presumed to be $\gg 1 $. Particularly,
$n_T=15$ for the two-valley electron
system in (100)
Si-MOS.
For low temperatures, $T\tau \ll 1$, in the so-called diffusive interaction regime,  $\Delta\sigma \propto
\ln(T\tau)$
depends
logarithmically on temperature;  for higher temperatures $T\tau\gg 1$, in the ballistic regime of interactions,
$\Delta\sigma$
varies linearly with $T\tau$. According to the IC theory, the crossover  occurs at $T_{\rm db} =
(1+F_0^\sigma)/2\pi\tau$
\cite{ZNA-R(T)_PRB_2001}, where
$F_0^\sigma$ is the Fermi-liquid coupling parameter.

Within the same approach, magnetoconductance in a weak in-plane magnetic field originates from  the field
dependence of the
effective number of  triplet channels, which in its turn is due to the Zeeman splitting mechanism
\cite{ZNA_BPar_2001}.
For example,
when the Zeeman energy $E_Z=g\mu_B B$ ($g=2$ is the bare g-factor for Si) becomes much greater than $T$ (but lower
than $T_F$), the
effective number of the triplet terms that contribute to $\Delta\sigma(T)$ is reduced from 15 to 7.

As a result, the first order interaction corrections to the MC  in the diffusive  and ballistic interaction regimes $\Delta\sigma
\equiv \sigma(T,B)-\sigma(T,0)$ may be written as follows \cite{ZNA_BPar_2001, vitkalov-valley_MR_PRB2003}:
\begin{eqnarray}
\Delta\sigma_d & \approx   A_d(F_0^\sigma,g_v) K_d(T,B,F_0^\sigma) \left(\frac{gB}{T}\right)^2, \quad  & T\tau \ll
1
\nonumber \\
\Delta\sigma_b &  \approx  A_b(F_0^\sigma,g_v) K_b(T,B,F_0^\sigma) (T\tau)\left(\frac{gB}{T}\right)^2, \quad &
T\tau
\gg 1.
\label{eq:MC_IC}
\end{eqnarray}
Explicit expressions for $K_b$ and $K_d$ are given in Ref.~\cite{ZNA_BPar_2001}.
In  terms of  Eq.~(\ref{eq:sigma}),   the above theory predictions are
 \begin{eqnarray}
 a_\sigma(T) \propto \left\{
\begin{array}{cl}
&(1/T)^2, \quad   T\tau \ll 1 \nonumber \\
&(1/T),    \quad  T\tau \gg 1.\\
\end{array}
\right.
  \end{eqnarray}

In Fig.~\ref{fig:Si4-rho(B2)}(a) the resistivity is somewhat lower than in Fig.~\ref{fig:Si4-rho(B2)}(b); this
difference is  due
to the different sample mobility.
It is worth mentioning that in the framework of the renormalization group theory \cite{finkelstein_1984,
punnoose_PRL_2002, pf_science_2005, castellani_PRB_1984, CCL_PRB_1998}, the magnetoconductance can also be
described
by the Castellani-–Di Castro-–Lee formula \cite{CCL_PRB_1998} which is equivalent to
Eq.~(\ref{eq:MC_IC})  in the diffusive limit and for $g\gg1$.

\subsection{High-  and low-mobility samples}
\begin{figure}[H]
\begin{center}
\includegraphics[width=0.49\linewidth]{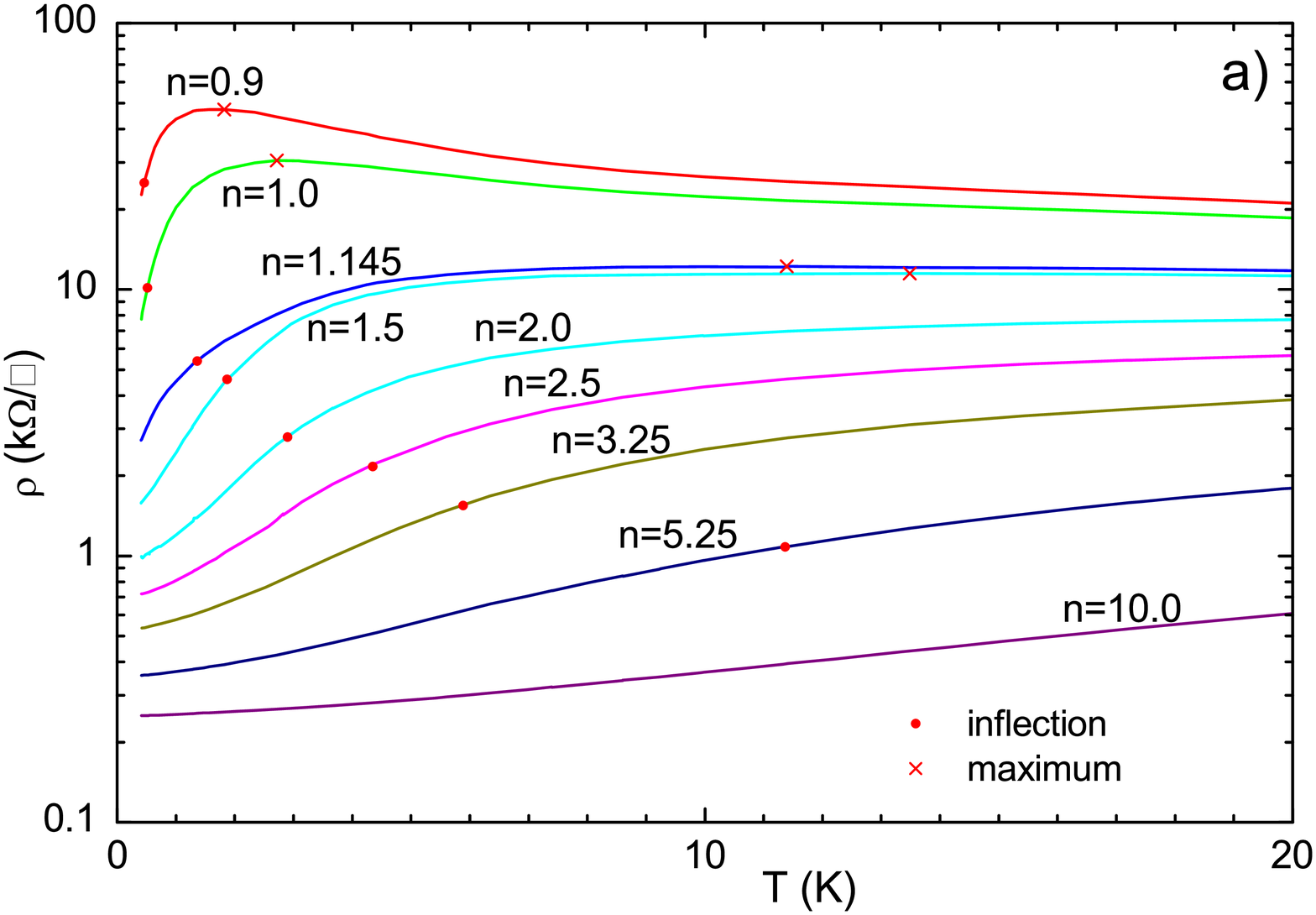}
\includegraphics[width=0.49\linewidth]{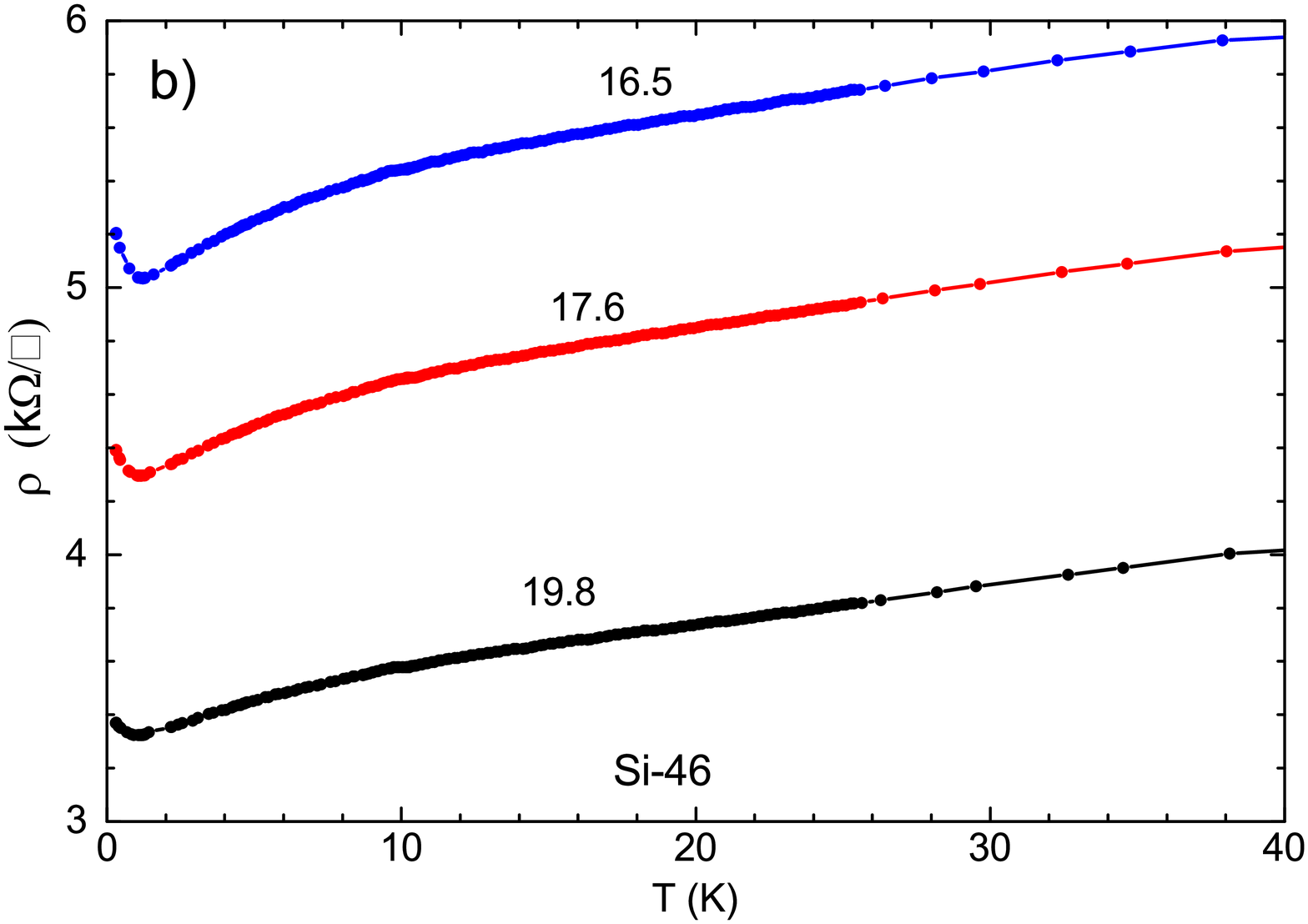}
\caption{(Color online) Temperature dependence of resistivity at zero field (a) for high mobility sample Si-2;
($n_c\approx 0.85$) at nine densities;  (b) for the low mobility sample Si-46 at three densities.
Carrier densities  are shown in units of $10^{11}$cm$^{-2}$.
Crosses and dots on the left panel mark the  $\rho(T)$ maxima, and the inflection points, respectively.
}
\label{fig:Si2-r(T)}
\end{center}
\end{figure}

Here, we  compare the magnetoconductivity behavior for  high- and low-mobility samples. At zero field,
the difference in their temperature dependencies  is illustrated in Figs.~\ref{fig:Si2-r(T)}(a) and
\ref{fig:Si2-r(T)}(b).
In the ``metallic'' range of densities, $n > n_c$, for the {\em high-mobility} samples Si-2, Si-4, Si-63, and
Si-6-14, the
resistivity
sharply varies by a factor of 6--10 \cite{krav_PRB_1995}. By contrast, for  the {\em low-mobility} samples Si-40
and
Si-46, $\rho(T)$ varies by $\approx
15\%$ only and its variation occurs at much higher temperatures and densities \cite{mauterndorf_1998}. These well
known features have been explored and
understood earlier \cite{mauterndorf_1998, klimov_PRB_2008, valley_scattering, proskuryakov_PRL_2002,
pudalov_R(T)_PRL_2003, pudalov_PSS_2004}.

In particular, the upturn at
low temperatures in Fig.~\ref{fig:Si2-r(T)}(b) is due to quantum corrections, which for low-mobility samples have
an
``insulating''
sign at all densities  (see Fig.~\ref{fig:Si2-r(T)}(b) and Ref.~\cite{mauterndorf_1998}). For high-mobility
samples,
the upturn sets
upon lowering  temperature,  ($T< 1/\tau; T <\tau_v$), where the effective number of triplet terms diminishes
\cite{klimov_PRB_2008},
and/or at  higher densities where $F_0^\sigma$ diminishes \cite{pudalov-logarithmic_JETPL_1998,
pudalov-Gmax_PRB_1999, pudalov_R(T)_PRL_2003, pudalov-gm_PRL_2002, clarke_NatPhys_2008}; these low-temperature and
high-density regimes  are out of sight in Fig.~\ref{fig:Si2-r(T)}.

 In the ``insulating'' regime,  the high- and low-mobility samples also have  distinctly different non-Ohmic and
 electric field  threshold conduction,
explored in detail in Refs.~\cite{pudalov_ECRYS_2002, pudalov_PRL_1993, pudalov-chui_PRB_1994,
pudalov-campbell_SurfSci_1996}.
These different features of the transport
result in a fundamentally different
behavior of  the correlation length: $\xi\propto \Delta/eE_t$ on the insulating side of the transition;
$\xi$ diverges as $ n\rightarrow n_c$ for high mobility samples, whereas  $\xi$ vanishes at $n_c$ for low-$\mu$
samples \cite{pudalov_ECRYS_2002, pudalov_PRL_1993}.

\begin{figure}[H]
\begin{center}
\includegraphics[width=0.5\linewidth]{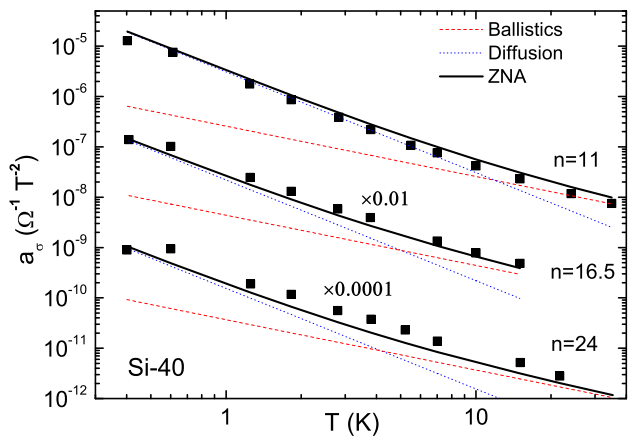}
\caption{(Color online) Temperature dependence of the $a_\sigma$ prefactor  for the low mobility sample Si-40
(filled
boxes). The densities  are indicated  in $10^{11}$cm$^{-2}$. The two higher density sets of data   are scaled by
the factors
indicated next to
each curve.
Dotted, dashed, and continuous bold lines show the predicted $a_\sigma(T)$ dependencies for ballistic,
diffusive and the total interaction correction, respectively, Eq.~(\ref{eq:MC_IC}).
}
\label{fig:a_sigma-Si40}
\end{center}
\end{figure}

The in-plane field magnetoconductance, which is the focus of our interest, for  low-mobility samples  develops in
accord with
interaction correction theory.
This is illustrated by Fig.~\ref{fig:a_sigma-Si40}, where the magnetoconductivity prefactor for  sample Si-40 is
shown versus temperature.
  The overall behavior is {\em quantitatively} consistent with the IC theory, Eq.~(\ref{eq:MC_IC}), which with no
  fitting parameters
  describes the low-temperature diffusive interaction   regime $a_\sigma \propto 1/T^2$, the high temperature
  ballistic  regime
  $a_\sigma \propto 1/T$, and the
diffusive-to-ballistic crossover at about $T=4- 5$\,K.

\begin{figure}[H]
\centering
\includegraphics[width=0.5\linewidth]{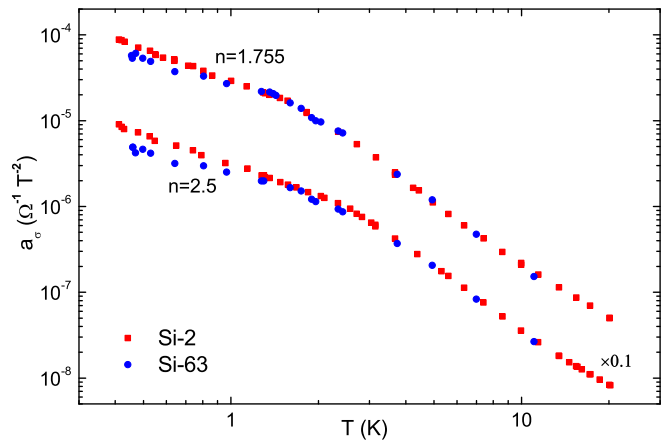}
\caption{(Color online) Comparison of the temperature dependencies of the  prefactors   $a_\sigma(T)$  for samples
Si2 and Si-63,
for two density values (indicated
in units of $10^{11}$cm$^{-2}$.  For clarity, the  curves are  scaled by the factors shown next to each curve.}
\label{fig:a_sigma-SiX}
\end{figure}

\begin{figure}[H]
\centering
\includegraphics[width=240pt]{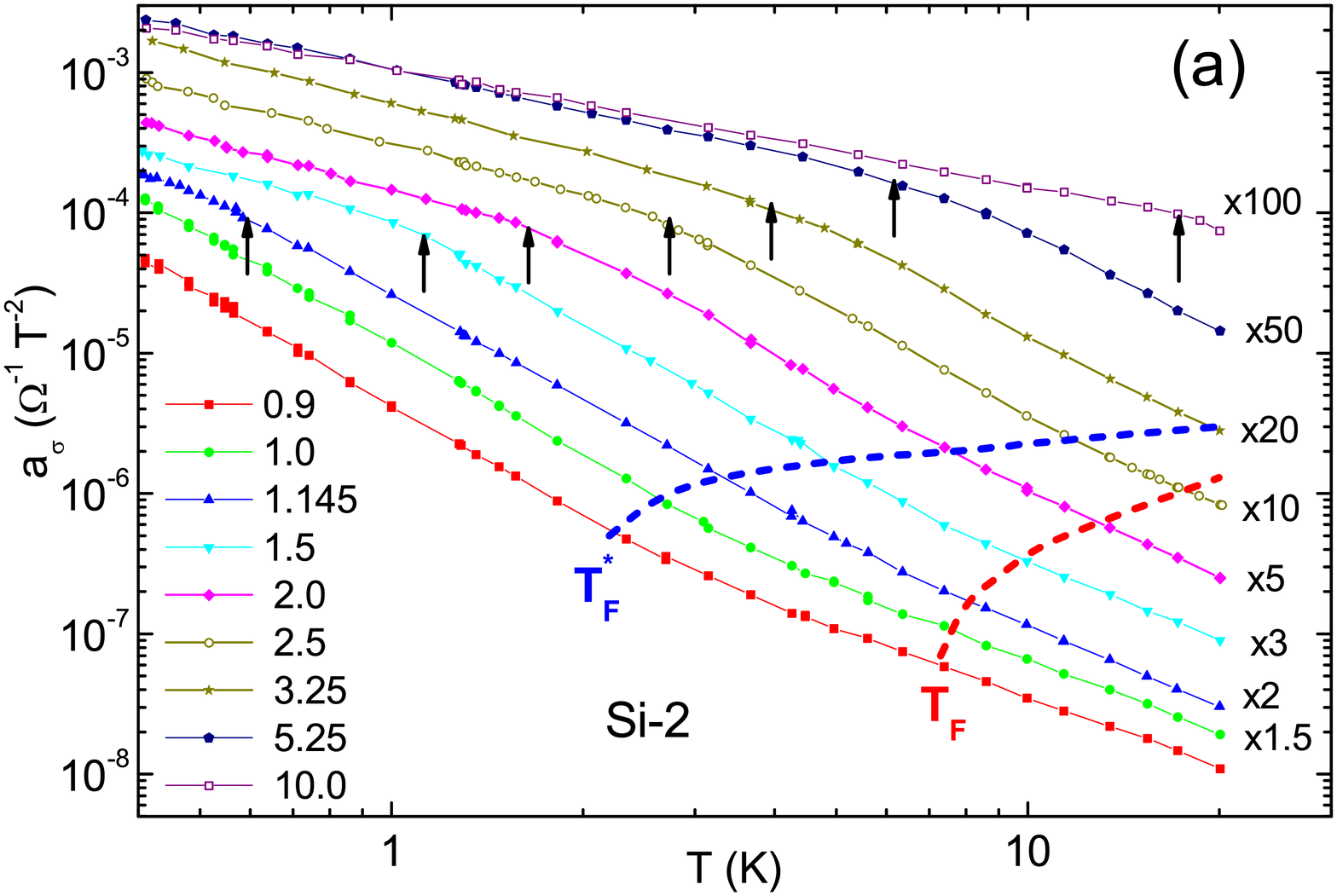}
\includegraphics[width=240pt]{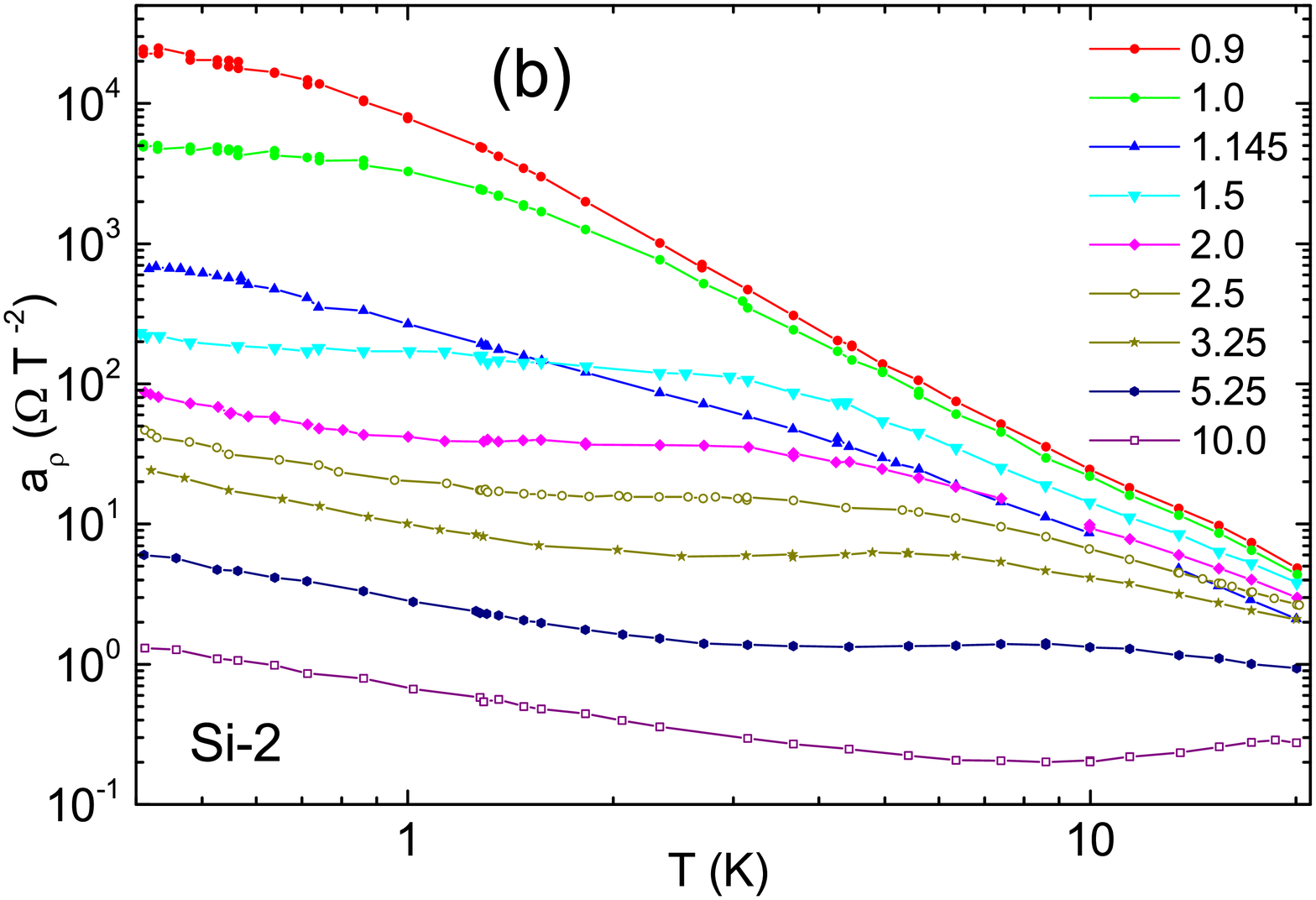}
\caption{(Color online) Temperature dependencies of the  prefactors  (a)  $a_\sigma(T)$ and (b) $a_\rho(T)$ for
sample
Si-2, for
several electron densities
indicated in units of $10^{11}$cm$^{-2}$.  On the panel a), for clarity, the  curves are  magnified by the factors
shown next to
each curve. Vertical arrows mark  the kink positions, the dashed
curves show $T_F(n)$ and $T_F^*(n)$.
}
\label{fig:a_sigma}
\end{figure}

The agreement with theory is no longer valid for the high-mobility samples.
In Fig.~\ref{fig:a_sigma-SiX}, we plotted the magnetoconductivity  prefactor
$a_\sigma(T,n)$ for the {\em high-mobility} sample versus temperature.
In this case, the estimated
diffusive/ballistic border
$T_{\rm db} \approx 0.2$\,K is below the accessible temperatures range of our
measurements and we anticipate to observe only the behavior characteristic of the ballistic regime.
One can see from Fig.~\ref{fig:a_sigma-SiX} that
 $a_\sigma(T)$  indeed develops in a ballistic fashion, $\propto T^{-1}$. This behavior extends up to temperatures
 1.5-2\,K (which
 is a factor of 10 higher than $T_{\rm db}\approx 0.2$K), then it sharply changes to  the
 unforseen  dependence,
 $a_\sigma(T)  \propto T^{-2}$, making the overall picture clearly inconsistent with theory predictions,
 Eq.~(\ref{eq:MC_IC}).  The  crossover in
Fig.~\ref{fig:a_sigma-SiX} occurs rather sharply,  as a kink on the double-logarithm scale.
The kink and the overall type of behavior  were  observed in a wide range of densities and were
qualitatively similar for the studied high-mobility samples (as  Fig.~\ref{fig:a_sigma} shows).

Figure \ref{fig:a_sigma}(a) shows the density evolution of  $a_\sigma(T)$ in a wide range of densities. Though the
high-temperature
behavior, $a_\sigma \propto T^{-1}$, formally coincides with that predicted for the diffusive interaction regime,
Eq.~(\ref{eq:MC_IC}), it extends up to temperatures of the order of Fermi temperature $T_F$. For this reason, this
behavior can not be associated with the diffusive interaction regime.

The immediate consequence of the $a_\sigma(T)$  behavior is that
the 2D electron system under study appears to have a  novel characteristic energy scale $T^* \approx
T_{\rm kink}(n)$.
The latter develops critically versus electron density, as  Fig.~\ref{fig:phase diagram} shows: $T_{\rm kink}$
vanishes
$\propto (n-n_c)$ at a finite density $n_c$, which is somewhat sample  dependent.
Within the experimental uncertainty,  this critical density for  $T_{\rm kink}(n)$ coincides with the MIT critical
density in transport \cite{krav_PRB_1995, knyazev_PRL_2008}.

The dependence  $T_{\rm kink} \propto (n-n_c)$  has little in common with the Fermi energy,  which in
the 2D-case  is
proportional to the carrier density $n$.
Clearly, the existence of such
an energy scale is inconsistent with  the Fermi liquid picture.
The critical $T_{\rm kink}(n)$
behavior points at the relevance of the electron-electron interaction effects.
Another indication of the crucial importance of the electron-electron interactions is the fact that the kink in
$a_\sigma(T)$ at
$T_{\rm kink}$ and the  anomalous regime of MC at $T>T_{\rm kink}$ are intrinsic only to high
mobility samples, where the strongly
correlated regime is accessed  upon lowering density. For samples Si-40 and Si-46 with a factor of 10 -- 30 lower
mobilities, such low densities are inaccessible and in
the same range of temperatures, the magnetoconductance  develops in accord with  IC theory predictions with no
kink.

The sharp crossover at high temperatures  to the  anomalous regime of MC, which is in
contrast with the theory predictions, is one of the main results of our study; it is intrinsic to high-mobility
samples and dilute regime of strong interactions.

\subsection{Magnetoconductivity and  magnetoresistivity}

For  high-mobility Si-MOSFETs and in the low-density and intermediate-temperature regime  ($1/\tau <T < T_F$)
the
in-plane field
magnetoconductivity  is inequivalent to the magnetoresistivity (MR), because variations of the conductivity
with temperature at zero field are large, a factor of 4 -- 10. As a result,  the $a_\sigma
(T)$ and $a_\rho(T)$ temperature dependencies are different  [see
Figs.~\ref{fig:a_sigma}(a) and \ref{fig:a_sigma}(b)]. The latter  is nonmonotonic and
less transparent, being affected by both, the onset of the
 anomalous regime in MC and by the strong $\rho(T)$ [and
$\sigma(T)$] variations.
For higher densities  $n=10, 5.25, 3.25\times 10^{11}$\,cm$^{-2}$, where the $\rho(T)$ variations are relatively
weak
[the lowest three curves in Fig.~\ref{fig:a_sigma}(b)],  $a_\rho(T)$
exhibits a shallow maximum that coincides with the kink in $a_\sigma(T)$.
For lower densities, the maxima in $a_\rho(T)$  get smeared, which hampers their quantification.
The simplicity of the $a_\sigma(T)$ dependence [in comparison with $a_\rho (T)$]
clearly points at the  primary role of the
{\em magnetoconductivity} rather than magnetoresistivity in the physical mechanism responsible for the appearance
of  the kink.

The kink temperature  $T_{\rm kink}$  lies far away from the bare and renormalized Fermi energy  and from the
crossover $T_{\rm db}=(1+F_0^a)/2\pi\tau \approx 0.2$\,K value \cite{ZNA-R(T)_PRB_2001}, which are the only known
energy scales in the Fermi liquid. We
interpret $T_{\rm kink}$  as a manifestation of an additional energy scale, beside the Fermi energy. Obviously, no
such energy scale may exist  in the pure 2D Fermi liquids, and vice versa, its existence  indicates a non-Fermi
liquid state.

 In Fig.~\ref{fig:a_sigma}(a), one can also see that the magnetoconductivity prefactor exhibits another twist
 upward
 for even  higher temperatures, clearly noticeable for the four  lowest curves (lowest densities). However, this
 feature occurs close to  the  renormalized $T_F$
and is likely to signify  a transition to a nondegenerate regime, which is beyond the scope of our paper.

\subsection{Other available data: spin magnetization}
In order to test whether the kink temperature  in magnetoconductivity has a more general  significance and indeed
signals a novel energy scale,
we inspected the temperature dependencies of other physical quantities measured in the  high temperature range
 and in weak or zero magnetic fields.
Available data that fit these requirements are as follows:
(i) spin magnetization per electron $\partial M/\partial n$ \cite{teneh_PRL_2012}, (ii) entropy per electron
$\partial S/\partial n$ \cite{kuntsevich_NatCom_2015}, and
 (iii) zero-field transport $\rho(T)$.

The spin magnetization-per-electron $\partial M/\partial n$  data \cite{teneh_PRL_2012}, in general,
 are interpreted as a clear evidence for the formation of a two-phase state, in which the Fermi liquid phase
 coexists with
 large--spin collective ``spin droplets'' (the latter being presumably collective localized states).
These data \cite{teneh_PRL_2012} show a pronounced sign change of $\partial \chi/\partial n \equiv
\partial^2 M/\partial B\partial n$ at a  density-dependent temperature $T_{dM/dn}(n)$.
Physically,  the sign change means that for temperatures   lower than $T_{dM/dn}(n)$, the minority phase (large
spin
collective ``spin droplets'') melt as density increases. In other words, extra electrons
added to the system join the Fermi sea, improve screening and favor ``spin droplets'' disappearance.
For temperatures above $T_{dM/dn}(n)$, the number of ``spin droplets''  grows as density increases; here the extra
electrons added to the 2D system prefer joining the ``spin droplets''.

\begin{figure}[H]
\centering
\includegraphics[width=0.6\linewidth]{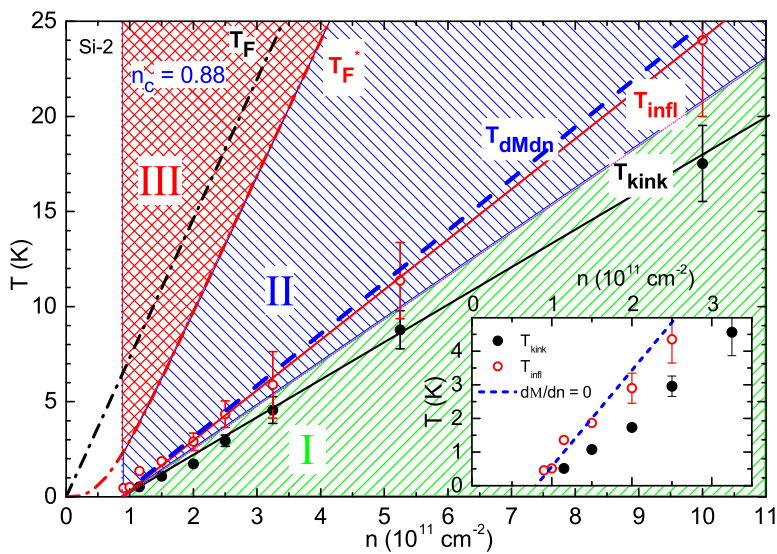}
\caption{(Color online) Empirical phase diagram of the 2DE system.
Dashed areas are: (I)
-- the ballistic interaction  regime, (II) --
the  anomalous MC regime. Hatched area (III) is the  nondegenerate regime, the blank area at $n < n_c$ is a
localized phase. Full dots:  the kink  temperature $T_{\rm kink}$; open dots:
the inflection point $T_{\rm infl}$.
Sample Si-2.  Dash-dotted curves show the calculated  bare ($T_F$) and the renormalized ($T_F^*$) Fermi
temperatures.
The insert blows up the low density region; the dashed line is $T_{dM/dn}$ \cite{teneh_PRL_2012}.
}
\label{fig:phase diagram}
\end{figure}
 The spin magnetization measurements \cite{teneh_PRL_2012} have been performed with our high mobility samples
 (almost
 identical   to   Si-2 and Si-63), and also with high-mobility Si-MOS samples from a different manufacturer
 \cite{reznikov_arxiv_2009}; all samples   demonstrated a universal   behavior. We believe therefore these results
 may be  compared with our current magnetotransport   data.
The  $T_{dM/dn}(n)$ dependence  copied from Figs.~1 and 2 of Ref.~\cite{teneh_PRL_2012} is depicted in the insert
to
Fig.~\ref{fig:phase diagram}.
One can see that $T_{dM/dn}(n)$ also behaves critically and vanishes to zero at $n_c$; remarkably,
within the measurements uncertainty, it is consistent with $T_{\rm kink}(n)$ deduced
from our  magnetotransport data.

With the same aim, we also inspect our earlier entropy-per-electron $dS/dn$ measurements
\cite{kuntsevich_NatCom_2015}. There is a clear onset of the strong $dS/dn$ growth with lowering density at
$n\approx n(T^*)$, signaling
a  crossover from the Fermi-liquid-type behavior  $dS/dn\approx 0$  to a  large entropy phase (see Figs.~1(a), and
1(c) of Ref.~\cite{kuntsevich_NatCom_2015}). The later phase corresponds  to the region II of the phase diagram in
Fig.~\ref{fig:phase diagram}. These data do not contradict the spin magnetization data and the empirical phase
diagram (Fig.~\ref{fig:phase diagram}), though do not enable us to explain  the magnetotransport puzzling behavior.
The latter will be done in the next sections.

\subsection{Other available data: resistivity and conductivity in zero field}
Searching for  manifestation of the  novel energy scale in zero-field transport,
we analyze the  $\rho(T)$ and $\sigma(T)$ dependencies at zero field (see Fig.~\ref{fig:Si2-r(T)}). The variations
of
these quantities in the relevant
temperature range are  large (up to a factor of  10), making the IC theory inapplicable in this
``high-temperature''
regime.

Each $\rho(T)$ curve has two remarkable  points:  $\rho(T)$ maximum, $T_{\rm max}$, and
inflection,  $T_{\rm infl}$ \cite{knyazev_PRL_2008}.  Whereas $T_{\rm max}$ is an order of the renormalized Fermi
energy, the  inflection point happens at much lower temperatures, in the degenerate regime.
Importantly, the inflection temperature
appears to be close to the kink temperature (see Figs.~\ref{fig:Si2-r(T)} and \ref{fig:phase diagram}). Therefore,
the
proximity of the three notable temperatures, which are inherent to high mobility samples solely,
$T_{\rm kink} \approx T_{\rm infl} \approx T_{dM/dn}$ strongly
supports  the existence of a new energy scale $T^*$ in the correlated 2D system.

$T^*$ is much less than
the bare Fermi temperature $T_F$ \cite{ando_review} and the renormalized  $T_F^* = T_F (m_b/m^*)$
\cite{pudalov-gm_PRL_2002}. In contrast to $T_F$ (which is $\propto n$), $T^*(n)$ develops
as ($n-n_c$).
On the other hand, $T^*(n)$  is much higher than   the ``incoherence'' temperature at
which the phase coherence is lost
(defined as $\tau_\varphi(T) =\tau$  \cite{brunthaler_PRL_2001}),
confirming that the kink, inflection, and $\partial \chi/\partial n$ sign change are irrelevant to the
single-particle
coherent effects.

\section{Discussion}
\subsection{Phenomenological model for transport and magnetotransport}
In the absence of an adequate microscopic theory, we attempt to elucidate the origin of the $T^*$ energy scale and
of the anomalous magnetoconductance behavior.
We suggest below a phenomenological  two-channel scattering  model that links the ``high temperature"
transport and magnetotransport behavior
in a unified picture  and makes a bridge to the thermodynamic magnetization data.  The physical picture behind the
two-channel scattering is described further, in the corresponding section.

One can see from  Fig.~\ref{fig:Si2-r(T)}  that the $\rho(T)$ temperature dependence is  monotonic up to the limits
of degeneracy,
$T=T_F$, and follows one and the same additive resistivity functional form  over a wide density range:
\begin{eqnarray}
\rho(T)&=&\rho_0 +\rho_1\exp(-\Delta(n)/T),\nonumber \\
 \Delta (n)&=& \alpha (n-n_c(B)),
\label{eq:rho-exp}
\end{eqnarray}
where  $\rho_1(n,B)$ is  a slowly decaying function of $n$, and $\rho_0(n,T)$ includes Drude resistivity and
quantum
corrections,
both from the single-particle interference and interaction \cite{remark-T-range}.
Although the above empirical  resistivity form has been suggested in Ref.~\cite{pudalov-SO_JETPL_1997} on a
different
footing, it fits well the $\rho(T)$
dependence for a number of  material systems \cite{pudalov-SO_JETPL_1997, hanein_1998, papadakis_1998, gao_2002,
brunthaler_SOI_2010, zhang-SrTiO3_PRB_2014, raghavan-SmTiO3_APL_2015, pouya-SrTiO3_PRB_2012}.

This empirical  additive $\rho(T)$ form  satisfies general requirements for the transport behavior in the vicinity
of
a critical
point \cite{amp_2001, knyazev_PRL_2008}, and explains the apparent success of the earlier attempts of
one-parameter
scaling [namely of the $\rho(T)$ steep rise and the mirror reflection symmetry between $\rho(T)$ and $\sigma(T)$ on
the metallic and insulating
sides of the MIT]  \cite{krav_PRB_1994, krav_PRB_1995}.
The additive resistivity form presumes the two-phase state of the
low-density 2D electronic system (cf. Matthiessen's rule). The two-phase state is experimentally  revealed in
macroscopic
magnetization measurements \cite{teneh_PRL_2012}, and in experiments with mesoscopic systems  or local probes
\cite{ilani_Science_2001, ghosh_PRL_2004}.
There is also a large body of theoretical suggestions for spontaneous formation of the two-phase state
\cite{spivak,
shi-xie_PRL_2002, sushkov, narozhny_2000, dharma_2003, khodel_2005, benenti_prl_2001, eisenberg_PRB_1999} due to
instabilities in the charge or spin channel.  Dealing with the two-phase state, the two channel scattering or
additive resistivity approach seems quite adequate to the problem.

The features of our interest, $T_{\rm kink}$ and $T_{\rm infl}$,
represent  ``high-energy'' physics.
Moreover, the  $\rho(T)$ [and $\sigma(T)$] variations of the experimental data
(Fig.~\ref{fig:Si2-r(T)}) are so large, that the first order in $T$ corrections, of cause, cannot describe them.
Our analysis of other known theoretical models for a homogeneous 2D Fermi liquid
 reveals that
neither of them describes  adequately the inflection on the $\rho(T)$ data and of course does not include an
associated energy scale. This is another motivation for us to turn attention to
the two-phase state.

The typical  $\rho(T)$ behavior   (Fig.~\ref{fig:Si2-r(T)}) naturally prompts the dual channel scattering.
The simplest functional dependence, Eq.~(\ref{eq:rho-exp}),   correctly describes  the inflection in $\rho(T)$ and
linear density
dependence of the inflection temperature   \cite{pudalov-SO_JETPL_1997, pudalov_disorder_2001, vitkalov-form}.
Obviously, in this model $T_{\rm infl}= \Delta/2$. To take magnetic field into account, and following results of
Refs.~\cite{pudalov_disorder_2001}
we include to $(\Delta/T)$ all the lowest order in $B/T$ (and even-in-$B$)  terms, as follows:
\begin{equation}
\Delta(T,B,n)/T =\Delta_0(n)/T - \beta(n) B^2/T  - \xi(n) B^2/T^2,
\label{eq:nc(B)}
\end{equation}
with $\Delta_0=\alpha[n-n_c(0)]$.

 Equations~(\ref{eq:rho-exp}) and (\ref{eq:nc(B)})
 link the magnetoconductance  with the zero-field $\rho(T)$
 temperature  dependence. With these, the $\rho(T,B)$ dependence is as follows:
\begin{eqnarray}
\rho(B,T) &=& \left[\sigma_D -
\delta\sigma \cdot \exp \left(- T/T_B \right)\right]^{-1} \nonumber \\
&+ &\rho_1 \exp \left( - \alpha \frac{n-n_c(0)}{T} - \beta \frac{B^2}{T} - \xi \frac{B^2}{T^2}\right)
\label{eq:r(B&n)}
\end{eqnarray}
The term in the square brackets includes the Drude conductivity and interaction quantum correction
\cite{ZNA-R(T)_PRB_2001,
ZNA_BPar_2001}. The latter, $\delta\sigma(T)= \gamma (B^2/T) + \eta T$,  was calculated using experimentally
determined $F_0^\sigma (n)$ values
\cite{pudalov-gm_PRL_2002, klimov_PRB_2008}, and $\sigma_D$ found from a standard procedure
\cite{pudalov_R(T)_PRL_2003}. In
order to cut off the corrections above a certain border temperature \cite{cut-off}  and, thus, to disentangle the
exponential- and linear-in-$T$  contributions,
the calculated interaction correction are cut-off with an exponential crossover function above $T_{\rm B} $, which
for
simplicity, we set equal to $\Delta(n)/2$.

\begin{figure}[H]
 \centering
\includegraphics[width=0.7\linewidth]{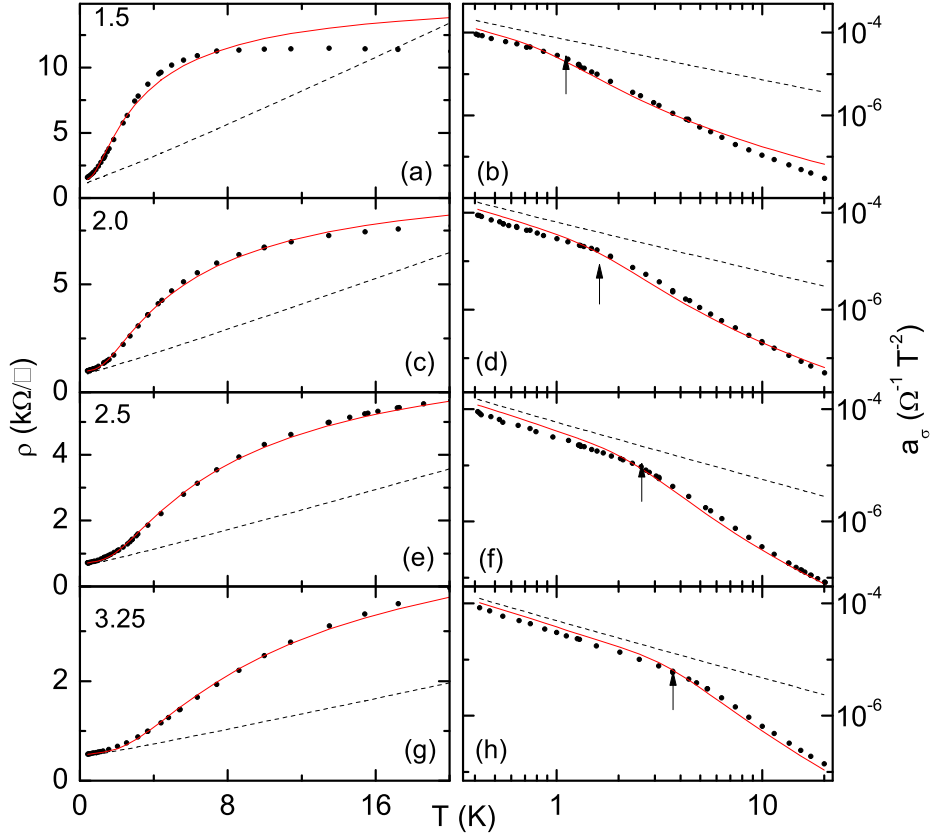}
\caption{(Color online) Fitting $\rho(T,B=0)$ dependencies (left) and  $a_\sigma(T)$ (right) with the same set of
the
fitting parameters. Sample
Si-2; carrier densities (from
top to bottom) are $n=1.5; 2.0; 2.5,$ and $3.25 \times 10^{11}$\,cm$^{-2}$. Fitting parameters are presented in
Table I. }
\label{fig:dualFit}
\end{figure}

From Eq.~(\ref{eq:r(B&n)}),
 the prefactor  $a_\sigma = - (1/2)\partial^2\sigma/\partial B^2 $ is calculated straightforward and
in Fig.~\ref{fig:dualFit}
is compared with experimental data.
In the $\rho(T)$ fitting [Figs.~\ref{fig:dualFit}(a), 7(c), 7(e), and 7(g)],  basically, there is only one
adjustable parameter, $\rho_1(n)$, for each density. Indeed,
$n_c(0)$ is determined from the conventional scaling analysis at $B=0$ \cite{knyazev_PRL_2008}, and the slope,
$\alpha = 2 \partial T_{\rm infl}(n)/\partial n$
may be determined  from Fig.~\ref{fig:phase diagram}. However, in order to test the assumed linear $\Delta(n)$
relationship, Eq.~(\ref{eq:rho-exp}), we treated  $\alpha(n)$ as an adjustable
parameter. On the next step, in the $a_\sigma(T)$ fitting [Figs.~\ref{fig:dualFit}(b), 7(d), 7(f), and 7(h)], we
fixed the parameters determined from the $\rho(T)$ fit, and varied  $\beta(n)$ and $\xi(n)$.

One can see that both $\rho(T)$ and $a_\sigma(T)$ are well fitted; the model captures correctly the major data
features, the steep $\rho(T)$  rise (including the
inflection), and the $a_\sigma(T)$ kink. Within this  model, the kink signifies a transition from the
low-temperature
magnetoconductance regime
[where  the linear $\sigma(T)$ temperature dependence
dominates and the exponential term may be neglected] to the high-temperature regime governed
by the steep exponential $\rho(T)$ rise; both regimes being irrelevant to diffusive interaction. The
parameters of the fit (Figs.~\ref{fig:dualFit}) are summarized in the Table I. The
factor $\beta$ is an order of magnitude smaller than $\xi$, therefore, the corresponding
term in Eq.~(\ref{eq:r(B&n)}) becomes important only at high temperatures. The slope, $\alpha$, is almost
constant,
confirming our assumption  [(Eq.~(\ref{eq:rho-exp})].

\begin{table}[h]
\caption{Summary of fitting parameters, corresponding to Fig.~\ref{fig:dualFit} and Eq.~(\ref{eq:r(B&n)}).
$\rho_1$
and $\rho_D =
\sigma_D^{-1}$  are in ($\Omega/\Box $), density is in units of $10^{11}$\,cm$^{-2}$,  $n_c=0.88$, $\alpha$ is in
K/$10^{11}$cm$^{-2}$.
}
\label{F0summarytable}
\begin{tabular}{|c|c|c|c|c|c|}
        \hline $n$ & $\rho_D$ & $\rho_1$ & $\alpha $ & $\beta$\,(K/T$^2$) & $\xi$\,(K$^2$/T$^2$)\\
		\hline 1.5 & 1268 & 14362 & 4.53 & -0.0160 & -0.08  \\
		\hline 1.996 & 901 & 9564 & 4.35 & -0.0080 & -0.09  \\
		\hline 2.5 & 662.2 & 6937 & 4.28 & -0.0043 & -0.11  \\
		\hline 3.25 & 501.5 & 5202 & 4.24 & -0.0019 & -0.15 \\
		\hline 5.252 & 336.14 & 3456.6 & 4.18 & -0.0005 & -0.19  \\
\hline
\end{tabular}
\end{table}

\subsection{Possible origin of the two channel scattering}
We suggested a unified phenomenological description of the transport and magnetotransport data, based on the
two-phase state (two scattering channels).
The two parallel dissipation channels in Eq.~(\ref{eq:rho-exp})
are (i) ordinary scattering (by impurities and interface roughness) of the itinerant electrons in 2D Fermi liquid,
and
(ii) Coulomb  scattering   of itinerant electrons by the charged collective localized states (``spin-droplets'').
The latter may be viewed as
quantum dots confining
four or more electrons \cite{teneh_PRL_2012}.
Besides the low-lying ground energy state, the dot (droplet)  contains an excited level, located above the Fermi
energy, at $E_F+\Delta$. Capture and emission of electrons from/to the surrounding Fermi
sea is a slow process, requiring rearrangement of all electrons inside the dot. Consequently, for a sufficiently
long time, much longer than the transport scattering time,
the dot may become charged and scatter itinerant electrons effectively. The probability of its charging is
negligible at low temperatures $T\ll \Delta$ but grows  with  temperature as $\exp(-\Delta/T)$.
The neutral dots (droplets)  do not scatter itinerant electrons because their size is larger than the Fermi wave
length.
As a result, the presence of droplets does not affect low-temperature transport, and magnetotransport at $T\ll
T^*(n)$.
Only at temperatures above $\Delta$ charging of droplets and
hybridization of itinerant and localized electrons  become significant and contribute to transport, leading to
the exponentially strong $\rho(T)$
temperature dependence  and the  anomalous
magnetotransport regime. This qualitative model is roughly similar to the charged trap model by Alltshuler and
Maslov \cite{am_traps}, but relates the traps with the spin droplets inside the 2D layer, rather than with defects
at the Si-SiO$_2$ interface.
In principle, the presence of the spin droplets is expected to cause saturation of the temperature dependence of
the phase breaking time  \cite{repin_PRB_2016}, however, we did not reveal the $\tau_\varphi$ saturation down to
about 30\,mK \cite{klimov_PRB_2008}; possible explanation of the low saturation temperature is discussed in
Ref.~\cite{repin_PRB_2016}.

\subsection{
On the magnetoconductivity interpretation}
For high densities $n\gg n_c$, the temperature  range above $T_{\rm kink}$
is unambiguously beyond the diffusive regime of interactions and, hence,  the  $B^2/T^2$ dependence is
the
high-temperature MC regime of the {\em non-diffusive type}.
Below $T_{\rm kink}$ the temperature is still higher than $T_{\rm db}$ and  the regime $a_\sigma \propto T^{-1}$
(see
Fig.~\ref{fig:a_sigma})
therefore is reminiscent of the  ordinary ballistic interaction regime \cite{ZNA_BPar_2001}. This conclusion is
confirmed by Figs.~\ref{fig:dualFit}
where the standard interaction corrections incorporated in Eq.~(\ref{eq:r(B&n)}) with experimentally determined
interaction parameters provide quite a successful fit below $T_{\rm kink}$.

The kink in Fig.~\ref{fig:a_sigma}(a) moves down as carrier density decreases. As a result,
  the $\delta\sigma \propto -(B^2/T^2)$ regime for low densities occupies more and more space and eventually,
  approaching $n=n_c$, extends down to the lowest temperature of our measurements, $T=0.3$K.
By tracing the evolution of this  regime from the higher-density side we conclude that this is  a high-$T$
phenomenon
that can hardly have diffusive interaction origin.
Therefore we conclude that
in the vicinity of the critical density, and at temperatures down to 0.3K, the MC is
governed by a high-temperature  mechanism
of an unknown origin. In other words, the MC in the vicinity of $n=n_c$
mimics the  behavior anticipated for the {\em diffusive regime} of electron-electron interaction
\cite{ZNA_BPar_2001, CCL_PRB_1998, knyazev_JETPL_2006}.

The temperature of the $\rho(T)$ maxima is even higher than $T_{\rm kink}\approx T_{\rm infl}$ and, hence, also
belongs to the high-temperature regime. This fact suggests that the $\rho(T)$ maximum is not caused by the
diffusive interactions, at least in the explored temperature range $T>0.3$\,K.

This finding requires to refine the RG  treatment of the experimental $\rho(T,B_\parallel)$ data in the
vicinity of MIT \cite{knyazev_JETPL_2006}, and particularly,
the phase diagram  of the 2D interacting and disordered systems deduced from fitting the experimental data within
this
approach \cite{punnoose_PRB_2010, anissimova_nphys_2007}. Indeed, in these studies, namely the $\rho(T)$-maximum
and the temperature dependence $(B^2/T^{2+\varepsilon})$ of the magnetoconductance (with $\varepsilon>0$)
were taken as  evidence for the diffusive interaction; the latter was used as an input to deduce the  temperature
renormalization of the interaction parameter $\gamma_2=-F_0^\sigma/(1+F_0^\sigma)\propto 1/T^\varepsilon$
\cite{knyazev_JETPL_2006}.
The new measurements of the in-plane field MR now should be taken at much lower temperatures, in the
millikelvin-range,  in
order to reveal the true diffusive regime for the high mobility samples
and to use these data  for comparison with the RG
theory~\cite{pf_science_2005}.  This however is experimentally challenging since requires  measurements of a tiny
magnetoresistance in extremely low fields $g\mu_B B <T$.

Our results also explain why the Fermi-liquid  parameters extracted from fitting the measured  magnetoconductance
scatter
significantly in various experiments and why they differ from those obtained from zero-field $\sigma(T)$ data:
indeed, by  fitting the data in the nominally ballistic regime,  one would
observe $a_\sigma$ (and deduce $F_0^a$ values) to be strongly dependent on
the particular temperature range, above or below the kink.

\section{Conclusion}
In conclusion, we have found
 unforeseen  features in transport and magnetotransport in the correlated 2D electron system
which set in above  a characteristic temperature $T^*$ that suggests a novel energy scale in the two-phase
electronic
system.
We attribute these features to the effect of spin-polarized collective droplets on transport and magnetotransport
of itinerant electrons in the correlated 2D electron system.
At the crossover  $T^*(n)$, the spin magnetization per electron changes sign, the in-plane field magnetoconuctance
crosses over from the conventional ballistic-type $-(B^2/T)$
 to the
  anomalous $-(B^2/T^2)$ dependence, and the zero field resistivity $\rho(T)$ exhibits an inflection, i.e.,
a transition from the linear-in-$T$ to the exponentially strong $T$-dependence.
The three respective temperature borders develop critically, $\propto (n-n_c)$, and are rather close to each other.
Since the
crossover at $T^*$ in the thermodynamic magnetization is related to the transition from
 growth to decay
of the SD phase, we conjecture that
$T^*$ might be related with the energy spectrum
of the spin droplets. The latter makes a bridge between the features observed in transport and thermodynamics.

We suggested a unified phenomenological description of the transport and magnetotransport data, based on the
two-phase state and two scattering channels. The two parallel dissipation channels in our models are presumably:
(i) ordinary
scattering of the itinerant electrons by impurities in 2D Fermi liquid, and (ii)  Coulomb  scattering (and,
possibly, hybridization) of itinerant
electrons by the collective localized states (``spin-droplets'').

Clearly, there is need for a microscopic theory that must link the transport and thermodynamic features and explain
on the same footing all three critical behaviors:
in the zero field resistivity, in the magnetoconductivity,
and in the spin susceptibility per electron. We
believe that an adequate theory should incorporate the two-phase state.
A possible key to the  origin of $T^*$  may be related with the  structure of the collective energy levels for the
individual
droplets of the minority phase, which in analogy with the quantum dots may simultaneously cause features  in
the thermodynamics and in the transport of itinerant electrons.

\subsection{APPENDIX: On the role of phonons}
In 3D metals  any residual weak temperature dependence in $\rho(T)$ originates from phonon
scattering which produces the  Bloch-Gruneisen behavior, $\rho(T)=\rho_0 +\rho_1 T^5$, where the temperature-
independent contribution $\rho_0$ arises from short-range disorder
scattering and the  temperature dependence (the second term) -- from  phonon scattering.  By contrast,
the temperature-dependent transport in 2D metallic systems at low temperatures, besides  weak-localization
effects,
 is dominated mostly by electron-impurity scattering dressed with electron-electron interaction effects (or on the
 complementary
 language -- by screened disorder scattering with temperature-dependent screening).

\begin{figure}[h]
\begin{center}
\includegraphics[width=0.8\linewidth]{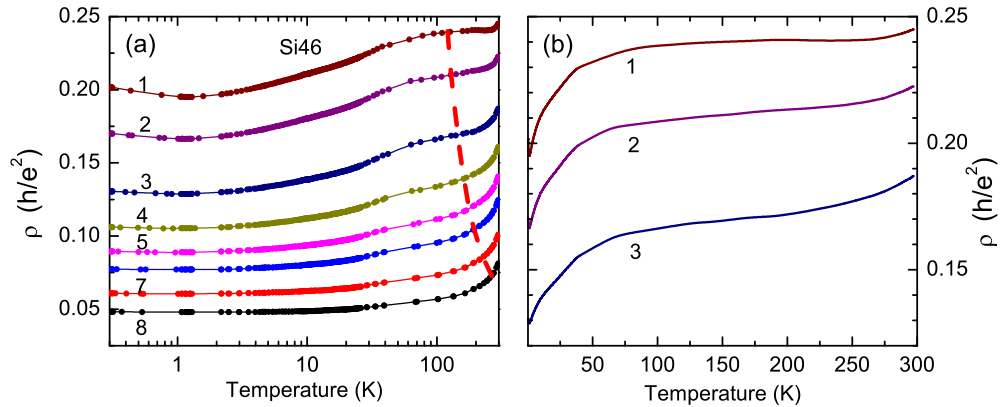}
\caption{(Color online)  Zero field temperature dependencies of  $\rho(T)$ for a low mobility sample in a wide
range
of
temperatures. (a)  logarithmic, and (b) linear scale. Sample Si-46. Density: 1 -- 16.5, 2 -- 17.6, 3 -- 19.8,
4 -- 22,
5 -- 24.2, 6
-- 26.4, 7 -- 30.8, and 8 -- 36.3 $\times 10^{11}$cm$^{-2}$. Dashed line depicts $T_F$ for various densities.}
\label{fig:phonons}
\end{center}
\end{figure}

 The interaction effects in transport  are proportional to $(T\tau)$ and in order to diminish them and to
 highlight
 the effect of
 phonons, we present in Fig.~\ref{fig:phonons} the resistivity data for low mobility Si-MOS sample (where ($\tau$
 is
 smaller by a  factor of 10 than for the high mobility samples studied in the paper). From
 Fig.~\ref{fig:phonons}(a) one can see  that below about 2\,K, logarithmic quantum corrections  dominate (both WL
 and interaction corrections)
\cite{amp_pss_2000}. For higher
temperatures, up to the Fermi energy (dashed curve), ballistic interaction corrections (or temperature-dependent
screening) take over and cause $\rho(T)$ growthe which flattens and then saturates as $T$ approaches $T_F$, due to
nondegeneracy effects \cite{dassarma_2000}.  For temperatures higher than 100-- 200\,K, resistivity again starts
growing, now due to electron-acoustic-phonon scattering. The monotonic $\rho \propto T$ dependence is a consequence
of the amount of
phonons excited at a given $T$.
In GaAs heterostructures, due to effective piezo-coupling, the phonon scattering is rather strong
\cite{dassarma_2000,
zhou_2012}.
For Si, the phonon scattering contribution to the overall scattering rate is much lower, because of the weaker
electron-phonon coupling mechanism (that is the deformation potential for Si).

 To conclude, it is well-known that phonon scattering in Si-structures contributes essentially to the transport
 only
 in the  vicinity of room temperature, and is irrelevant to the low-temperature transport. Both the nondegeneracy
 and phonon
 scattering  are  irrelevant to the inflection in $\rho(T)$ which happens at much lower temperatures than the onset
 of phonon
 scattering.
Nevertheless, to be on the safe side,  in our studies, we analyze the data [kink in $\partial^2\sigma/\partial B^2$
and
inflection
in $\rho(T)$] only in the temperature range (i) well below $E_F$  and below the $\rho(T)$ maximum in order to
avoid
the
nondegeneracy effects, and (ii) always below 20K for the explored densities, where the  phonon contribution to the
resistivity in Si-MOSFETs can be neglected with  1\% or better accuracy.

\section{Acknowledgements}
We thank I.\,S.~Burmistrov, I.~Gornyi, and A.\,M.~Finkel'stein for discussions. The magnetotransport  measurements
were supported by RFBR (15-02-07715, and 14-02-31697),
transport and thermodynamic measurements - by Russian
Science Foundation (14-12-00879). The work was done using research equipment of the Shared Facilities Center at
LPI.


\begin{thebibliography}{99}
\bibitem{remark-T-range} We restrict our considereation to the  ``intermediate'' and ``high-temperature'' regimes
$T_F > T> 1/\tau$ and neglect the logarithmic type dependencies which are isignificant in this temperature range.
\end{thebibliography}
\end{document}